\documentclass[final,rtqno]{siamltex}
\usepackage{stmaryrd}
\usepackage{amsmath}
\usepackage{amssymb}
\usepackage{multirow}
\usepackage{cite}
\usepackage{color}
\usepackage{comment}
\usepackage{tabularx}
\usepackage{algorithm}
\usepackage{algorithmic}
\usepackage{tabularx,epsfig,color,tabularx}
\usepackage{algorithm,algorithmic,multirow}

\input psfig.sty

\newcommand\bn{\textbf{n}}
\newcommand\bm{\textbf{m}}

\newcommand\bR{\mathbb{R}}
\newcommand\bau{{\textbf B}_1}

\newcommand{\be}{\begin{equation}}
\newcommand{\ee}{\end{equation}}
\newcommand{\ba}{\begin{array}}
\newcommand{\ea}{\end{array}}
\newcommand{\bea}{\begin{eqnarray}}
\newcommand{\eea}{\end{eqnarray}}
\newcommand{\beas}{\begin{eqnarray*}}
\newcommand{\eeas}{\end{eqnarray*}}
\newcommand{\bx}{{\bf x}}
\newcommand{\by}{{\bf y}}
\newcommand{\bz}{{\bf z}}
\newcommand{\bzo}{{\bf 0}}
\newcommand{\bk}{{\bf k}}
\newcommand{\erh}{E}

\newcommand*\samethanks[1][\value{footnote}]{\footnotemark[#1]}
\renewcommand{\theequation}{\arabic{section}.\arabic{equation}} %
\newtheorem{exmp}{Example}
\newtheorem{remark}{Remark}[section]
\newtheorem{prop}{Proposition}

\title{Accurate and efficient computation of nonlocal potentials based on Gaussian-sum approximation}
\author{Lukas Exl\thanks{Fak. Mathematik, Univ. Wien, Oskar-Morgenstern-Platz 1, 1090, Vienna, Austria ({\tt lukas.exl@univie.ac.at}). 
Inst. of Solid State Physics, TU Wien, Karlsplatz 13, 1040, Vienna, Austria.} 
\and Norbert J. Mauser\thanks{WPI c/o Fak. Mathematik, Univ. Wien,
 Oskar-Morgenstern-Platz 1, 1090, Vienna, Austria. (\tt norbert.mauser@univie.ac.at, yong.zhang@univie.ac.at)} \and Yong Zhang \samethanks }

\begin{document}

\maketitle


%
\begin{abstract}
We introduce an accurate and efficient method for a class of nonlocal potential evaluations with free boundary condition, including the 3D/2D Coulomb, 2D Poisson and 3D dipolar potentials. Our method is based on a Gaussian-sum 
approximation of the singular convolution kernel and Taylor expansion of the density. 
Starting from the convolution formulation, for smooth and fast decaying densities, we make a full use  of 
the Fourier pseudospectral (plane wave) approximation of the density and a separable Gaussian-sum approximation of the kernel in an interval where the singularity (the origin) is excluded. 
Hence, the potential is separated into a regular integral and a near-field singular correction integral, where the first
integral is computed with the Fourier pseudospectral method and the latter singular one can be well resolved utilizing 
a low-order Taylor expansion of the density. Both evaluations can be accelerated by fast Fourier transforms (FFT).
The new method is accurate (14-16 digits), efficient ($O(N\log N)$ complexity), low in storage, easily adaptable to other different kernels, applicable for anisotropic densities and  highly parallelable.
\end{abstract}


\pagestyle{myheadings}\thispagestyle{plain}
 \markboth{L.Exl, N.J. Mauser and Y. Zhang}
{ Gaussian-sum based nonlocal potential solver}

\section{Introduction}
In this paper, we aim to evaluate nonlocal potentials given originally by convolutions:
 \bea\label{ConvGen}
u (\bx )  = (U\ast \rho) (\bx)= \int_{\mathbb R^d} U(\bx-\by) \rho(\by) d\by,  \quad \quad   \bx \in \mathbb R^d, \quad d= 2,3,
\eea
where $\ast$ denotes the convolution operator, $\rho(\bx)$ is the density function, 
and $U(\bx)$ is a nonlocal (long-range) convolution kernel. Here we are interested in the cases where the density is smooth and decays fast.

Nonlocal potentials exist in a variety of mathematical models from quantum physics / chemistry to material sciences, plasma physics and computational biology etc.  The computation of the nonlocal potential is often the most 
time-consuming part in simulations, and the development of accurate and efficient numerical schemes remains an active and important topic
in the science and engineering community. The 3D Coulomb potential (also called "Newtonian potential"), with  $U(\bx) = \frac{1}{4\pi|\bx|}$, is fundamental and universal in many applications, such as Bose-Einstein Condensates \cite{BC7,DipJCP,Dip-NUFFT,YY2001,YY2000} and quantum chemistry\cite{PlaneGauss,Wavelet06,Wavelet07,Ewald}. 
Here, we study a class of nonlocal potentials with their kernels given explicitly as follows:
\be\label{kernelPhy}
 U(\bx)=\left\{\ba{ll}
\frac{1}{4\pi|\bx|}, & 3D \text{ Coulomb} ,\\[0.5em]
\frac{1}{2\pi|\bx|}, & 2D \text{ Coulomb} ,\\[0.5em]
-\frac{1}{2\pi}\ln |\bx|, &2D \text{ Poisson}.\\[0.5em]
\ea\right. 
 \ee

Of course, there are other nonlocal potentials and some of them can be reformulated through the above kernels, e.g., the dipolar convolution kernel $U(\bx)=\frac{3}{4\pi}\frac{\bm\cdot\bn-3(\bx\cdot \bn)
(\bm\cdot\bx)/|\bx|^2}{|\bx|^3}$ where $\bn,\bm \in \mathbb R^3$ are unit vectors \cite{BC7,DipJCP,Dip-NUFFT,BaoJiangLeslie,YY2000,YY2001}.
We restrict ourselves to those given in \eqref{kernelPhy} because they are both common and important.

The convolution \eqref{ConvGen} can be represented formally as a Fourier integral 
\bea\label{FouriGen}
u(\bx) = \frac{1}{(2\pi)^d}\int_{\mathbb R^d} \widehat{U}(\bk)  \widehat{\rho}(\bk) \,e^{i \bk \cdot \bx} d \bk, \qquad \bx \in \mathbb R^d,
\eea
where $\widehat{f}(\bk) =  \int_{{\mathbb R}^d} f(\bx)\;e^{-i \bk\cdot \bx}\, d\bx$
is the Fourier transform of $f(\bx)$ for $\bx,\bk\in \mathbb{R}^d$. Note that the Fourier transform of the
convolution kernel $\widehat{U}(\bk)$ is also long-range and singular, and sometimes the singularity is too strong that the Fourier representation is not well-defined, e.g., $1/|\bk|^2$ for the 2D Poisson potential \cite{SP-NUFFT}.  
Another important equivalent formulation is to solve a partial differential equation in the whole space with appropriate far-field condition.
For example, the 3D Coulomb potential satisfies the following Poisson equation, i.e.,
\begin{equation}\label{PoiDiff}
-\Delta\, u(\bx) = \rho(\bx), \qquad \bx \in \mathbb R^3, \quad
\qquad \lim_{|\bx|\to\infty}u(\bx)=0.
\end{equation}

All of the three formulations are challenging numerically. 
In \eqref{ConvGen} and \eqref{FouriGen}, the convolution kernels and their Fourier transforms are long-range and singular at the origin and/or at the far-field. Therefore, 
accurate and efficient evaluation requires either a large computational domain and/or elaborate strategies to take care of the singularity in physical and phase space, respectively.

Various numerical methods have been proposed to solve the potential via the PDE approach on a rectangular domain with uniform 
mesh grid\cite{DipJCP,BJNY,SPMCompare}. 
Take the 3D/2D Coulomb potential as an example. As the potential decays to zero at the far-field, the commonly used periodic and homogeneous boundary conditions, imposed on the boundary of the rectangular domain, do not agree very well with the far-field asymptotics. Errors coming from the boundary condition approximation dominates as the mesh size tends smaller. 
The saturated accuracy achieved by Fourier/Sine pseudospectral methods, also referred to as ``locking" accuracy, improves when the domain size increases \cite{BC7,DipJCP,SP-NUFFT,Dip-NUFFT}. 
While for the 2D Poisson potential, periodic or homogeneous boundary condition approximation
is totally inappropriate, 
because the potential diverges,  i.e., $u(\bx) \rightarrow C \ln |\bx|, C>0$  as $\bx \rightarrow \infty$. 
For this case, exact artificial boundary conditions on a disk were given by Zhang and Mauser\cite{ExtBd}, and boundary conditions on the rectangular domain remain to be further explored. 
However, not all interesting potentials can be formulated via PDE \cite{BJNY,SP-NUFFT}, therefore, we start from the convolution or the Fourier integral.   

Starting from the Fourier integral \eqref{FouriGen}, simple plane-wave discretizations suffer serious accuracy loss
due to improper treatment of the singularity in $\widehat{U}$ \cite{SP-NUFFT,Dip-NUFFT,PlaneGauss}. For kernels with removable singularity in spherical/polar coordinates, e.g., 
the 3D/2D Coulomb potential, Jiang, Greengard and Bao\cite{BaoJiangLeslie} proposed an accurate and efficient method by splitting the kernel into long-range regular and short-range singular part, and evaluating the quadrature 
via Fast Fourier Transform (FFT) and the nonuniform FFT (NUFFT) \cite{nufft2}, respectively.
This approach was recently adapted to the 2D Poisson potential case \cite{SP-NUFFT}, whose singularity is too strong to be cancelled out in polar coordinates. 
Their method can achieve spectral accuracy with great efficiency that is inherited from the FFT and NUFFT algorithm. However, it is not ideal because of the large prefactor in front of the $O(N\ln N)$ coming from the NUFFT\cite{nufft2,BaoJiangLeslie}. 
More importantly, the 3D Coulomb/dipolar evaluation is rather slow and needs further 
investigation for potential simulations. 

It is more natural to start from the convolution form \eqref{ConvGen}. In fact, there has been lot of work on this problem, see \cite{fast_conv_beylkin,Wavelet3,PlaneGauss,Wavelet06,Wavelet07}. 
A basic idea is to modify the kernel somehow and to evaluate the long-range interaction efficiently. 
Several methods have been proposed, such as the Ewald-type partition\cite{Ewald}, kernel-truncation \cite{Wavelet3}, Gaussian-sum (GS) approximation \cite{fast_conv_beylkin,Wavelet06,Wavelet07} etc. 
Among these approaches, the Gaussian-sum based method is one of the most effective and accurate solvers. The Gaussian-sum approximation has been studied intensively, we refer the readers to \cite{KernelGauss1,KernelGauss,Braess_expsum,steger_1993}.
In \cite{fast_conv_beylkin}, Beylkin et al. split the 2D/3D Helmholtz potential into a convolution with a Gaussian-sum in the spatial domain and band-limited multiplier in the Fourier domain. 
Later, Genovese et al.\cite{Wavelet06} solved the Poisson potential by combining interpolating scaling function (ISF) representation of the density and the Gaussian-sum approximation. The resulting discrete convolution was accelerated by FFT and it is ideal for parallelization. 
However, the optimal accuracy, around 10-digits, is limited by the resolution of the kernel's GS approximation and also the neglected near-field correction integral. 
 
Here we aim to design a method to combine the advantages of the NUFFT
and the ISF Poisson solver. To this end, we shall adopt the Gaussian-sum approximation for the regular long-range 
regular integral and  compute the near-field correction integral with local interpolations instead of global spectral interpolation, as in \cite{BaoJiangLeslie}. 

We first use a separable approximation to split the potential into a long-range regular and a short-range singular integral. The separable approximation is chosen as a Gaussian-sum and computed by 
sinc quadrature \cite{steger_1993} within a very high-resolution, i.e., about $10^{-16}$ in relative norm, over an interval $[\delta,2]$ with a relative large $\delta=10^{-3}$ or $10^{-4}$. 
This grants us the possibility to restrict the local correction in the interval $[0,\delta]$. We mention that this approach has already been proven effective by Exl and Schrefl \cite{exl_2014_nfft} in the context of computational micromagnetics.
Plugging the finite Fourier series expansion into the Gaussian convolution,
the evaluation boils down to four Fourier transforms (forward and backward pair counted as two). 
The sinc approach gives us a suitable, fast and easily adaptable way to obtain a Gaussian-sum approximation on a given interval within a prescribed accuracy. However, we shall remark that the sinc approach does not lead to an optimal approximation in terms of a minimal amount of Gaussian terms. 
Best approximation of singular kernels by means of numerical optimization was considered and successfully applied in \cite{Braess_expsum,  Wavelet3, exl_2014_nfft,HackBush2005}.  

In the computation of the regular integral, the tensor product structure of the Gaussian-sum approximation is exploited for accurate and stable pre-computation of the coefficients, which are given by higher-dimensional integrals.
In practice, the potential has to be solved many times, so it is worthwhile storing the precomputed coefficients so as to save storage and CPU-time. The near-field correction integration over the small ball $\mathcal B_\delta :=\{\bx\in \mathbb R^d\big| |\bx|<\delta \}$ is computed based on a low-order Taylor expansion of the density. Similarly,
the derivatives involved are also computed via FFT, therefore the near-field computation is also suitable for parallelization. We validate our approach for different types of nonlocal potentials, i.e., the 3D/2D Coulomb potential, the 2D Poisson potential and the 3D dipolar potential. 
Furthermore, an adaption to Davey-Stewartson nonlocal potential \cite{DS_StimmingZhang} is presented.

The paper is organized as follows.
In Section 2, we describe the algorithm, which consists of three steps: reformulation, long-range regular integral evaluation, short-range singular integral evaluation, followed by extensions to 3D dipolar potential. Detailed error analysis is also given. 
In Section 3, we present details on the Gaussian-sum approximation of two kernels, i.e., the Coulomb kernel $1/r$ and 2D Poisson kernel $\ln r$, by sinc quadrature. 
In Section 4, extensive numerical results are given to illustrate its performance in both accuracy and efficiency.
Finally, some concluding remarks are drawn in Section 5.

\section{Numerical algorithm}\label{method}
For the sake of consistency, we first rescale the problem  and introduce the reformulation of \eqref{ConvGen}. The key formulation is given in \eqref{key_form}-\eqref{key_form2}. 
The subsequent subsections explain the computation of the regular integral and the correction integral, as well as extension to 3D dipolar potential and application to 2D/3D Coulomb potential with anisotropic densities.

\subsection{Preliminary discussion}\label{prelim}
As assumed in the beginning that the density is smooth and fast decaying, we can reasonably assume that the density is compactly supported in a square box ${\textbf B}_L := [-L,L]^d \subset \mathbb R^d,d=2,3$ with a controllable precision. For the sake of simplicity, we choose a square box ${\textbf B}_L$, and 
it will be shown in the forthcoming subsection that a general rectangular box is also feasible.
The domain ${\textbf B}_L$ is also the domain of interest for the nonlocal potential $u$ \eqref{ConvGen}.

Following  a standard scaling argument, we first rescale the density to be compactly supported in the unit box $\bau$, i.e., 
\bea\label{rescale}
\bx = \widetilde{\bx} \;L, \quad \rho(\bx) = \widetilde{\rho}(\widetilde{\bx}) ,
\quad  \Longrightarrow \quad \widetilde{\bx} \in \bau,\;\; \text{supp}(\widetilde{\rho}) \subset \bau.
\eea
Plugging \eqref{rescale} into the convolution \eqref{ConvGen}, we have 
\bea
u(\bx)  = \int_{\mathbb R^d} U(\bx-\by)\rho(\by) d\by = \int_{{\textbf B}_L} U(\bx-\by)\rho(\by) d\by  = L^d \int_{{\textbf B}_1} \widetilde{U}(\widetilde{\bx}-\widetilde{\by})\widetilde{\rho}(\widetilde{\by}) d\widetilde{\by}.\quad \quad 
\eea
Particularly, for the 2D/3D Coulomb potentials, we have $\widetilde{U}(\widetilde{\bx}) = U(\bx) = U(\widetilde{\bx}L ) = L^{-1} U(\widetilde{\bx})$, therefore, 
\bea
u(\bx) =\widetilde{u}(\widetilde{\bx}) =  L^{d-1} \int_{{\textbf B}_1} U(\widetilde{\bx}-\widetilde{\by})\widetilde{\rho}(\widetilde{\by}) d\widetilde{\by}, \qquad \widetilde{\bx}  \in {\textbf B}_1, \quad d = 2,3. \quad \quad 
\eea
Similarly, for the 2D Poisson potential, we have
\bea
u(\bx) =\widetilde{u}(\widetilde{\bx}) =-\frac{L^2}{2\pi} \int_{ {\textbf B}_1} \widetilde{\rho}(\widetilde{\by})\,\ln{|\widetilde{\bx}-\widetilde{\by}|} \, {d} \widetilde{\by} - \frac{L^2}{2\pi}\ln{L} \int_{{\textbf B}_1} \widetilde{\rho}(\widetilde{\by}) \; {d} \widetilde{\by}, \quad \widetilde{\bx} \in {\textbf B}_1.
\eea
Notice that the domain of interest is also rescaled to the unit box ${\textbf B}_1$, therefore, the evaluation of $u(\bx)$ on ${\textbf B}_L$ is equivalent to computing $\widetilde{u}(\widetilde{\bx})$ on the unit box with rescaled density $\widetilde{\rho}(\widetilde{\bx})$, which is also compactly supported in ${\textbf B}_1$. We shall omit $\,\widetilde{}\,$  hereafter for simplicity. In practice, the computation domain ${\textbf B}_1$ is usually discretized uniformly in each direction, and the density is given on the uniform grids $\mathcal{T}_h$:
\begin{align*}
\mathcal{T}_h = \{(x_{j_1}^{(1)}, \hdots, x_{j_d}^{(d)})\big|\,x_{j_p}^{(p)} = -1 + j_p h^{(p)}, h^{(p)} = 2/n_p, 1\le j_p \le n_p,\, 
p = 1,\ldots,d\}.
\end{align*}

As discussed earlier, one of the key ideas is to use a separable GS approximation of the convolution, to be elaborated in Sec.~\ref{1_r} and \ref{logr}, to reformulate the potential into two integrals, namely, the long-range regular integral 
and short-range singular integral. To be precise, we can reformulate the potential \eqref{ConvGen} as follows:
\bea\label{key_form}
u(\bx) &=& \int_{\bR^d}U_{GS}(\by)\; \rho(\bx-\by) {d} \by +  \int_{\bR^d}\big( U(\by)-U_{GS}(\by) \big)\; \rho(\bx-\by) {d}\by \\
\label{key_form2} &:= & I_1(\bx)  + I_2(\bx), \quad \qquad \bx \in {\textbf B}_1,
\eea
where  $ U_{GS}$, the GS approximation of the kernel, is given explicitly as follows:
\bea\label{GS-Gene} U_{GS}(\by)= U_{GS}(|\by|):=\sum_{q= 0}^S w_q\, e^{-\tau_q^2 |\by|^2}, \quad S\in \mathbb N^{+}, \eea
with weights and nodes $\{(w_q, \tau_q)\}_{q=0}^{S}$, $I_1(\bx) $ and $I_2(\bx) $ are the long-range regular integral and short-range singular integral (also named as \textit{correction integral}), respectively. 
It is noteworthy to point out that the singularity of the integrand of $I_2$ at the origin in physical space is cancelled out in spherical coordinates by the Jacobian of the coordinates transform. 

Another important feature is our high-resolution GS approximation of the singular kernel $U$ over a interval excluding the origin $r=0$. In fact, with sinc quadrature\cite{exl_2014,steger_1993}, the GS approximation error $\varepsilon$, measured in relative/absolute maximum norm, over the interval $[\delta, 2]$ can be achieved up to machine precision. 
In our algorithm, the parameter $\delta$ does not have to be chosen as small as in \cite{Wavelet06} and we can choose some intermediate value, e.g., $10^{-4},10^{-3}$. With accurate GS approximation, the integral domain of the correction integral can be compressed to a small neighbourhood of the origin, i.e., ${\mathcal B}_\delta$. Thus 
the correction integral evaluation can be done with some Taylor expansion. Details are to be presented in subsection
\ref{corr}. Detailed description of the evaluation of $I_1$ and $I_2$ is to be shown in subsection \ref{I1} and \ref{corr}, respectively. 

\subsection{Evaluation of the regular integral $I_1(\bx)$}\label{I1}
Due to the compact support assumption of the density, plugging the explicit GS approximation \eqref{GS-Gene} into $I_1(\bx)$, we have 
\bea
\label{I1InteSt1} I_1(\bx) &= &  \int_{\bR^d} \sum_{q = 0}^S w_q\, e^{-\tau_q^2 |\by|^2} \rho(\bx-\by) {d} \by , \qquad \bx \in \bau,\\
\label{I1InteSt2}&=&\sum_{q= 0}^S w_q \int_{{\textbf B}_{\bx,1}} e^{-\tau_q^2 |\by|^2} \rho(\bx-\by) {d} \by,\quad \;\;\bx \in \bau,\\
\label{I1InteSt3} &=& \sum_{q = 0}^S w_q \int_{{\textbf B}_2} e^{-\tau_q^2 |\by|^2} \rho(\bx-\by) {d} \by, \qquad \bx \in \bau,
\eea
where ${\textbf B}_{\bx,1} := \textbf{B}_1+\bx$ is the unit box centred at $\textbf x$. Identity \eqref{I1InteSt3} holds 
because $\rho(\bx-\by)=0,\forall\; \bx \in \textbf{B}_1,  \;\by \in  \textbf{B}_2\setminus {\textbf B}_{\bx,1}$.
For $\bx \in \bau$ and $\by \in {\textbf B}_2$ holds $\bx-\by \in {\textbf  B}_3$, and we can approximate the density on ${\textbf B}_3$ by Fourier pseudo-spectral method with spectral accuracy \cite{Spectral}. 
To be more specific, a simple zero-padding of the density from ${\textbf B}_1$ to ${\textbf B}_3$ is applied first, and the padded density $\rho$ is well resolved by the following finite Fourier series:
\bea\label{FourSeriB3}
\rho(\bz) \approx \sum_{\bk} \widehat{\rho}_\bk\;  \prod_{j = 1}^d  e^{\frac{\;2\pi i \;k_j}  {b_j-a_j} (z^{(j)} - a_j)}, \quad \quad \bz = (z^{(1)},\hdots,z^{(d)}) \in {\textbf B}_3,
\eea
where $a_j=-3,b_j =3, j = 1\hdots d$ and $\bk = (k_1,\hdots,k_d) \in \mathbb Z^d$ with $k_j =-\widetilde{n_j}/2, \ldots,\widetilde{n_j}/2-1$ and $\widetilde{n_j} = 3 n_j$.
The Fourier coefficients are determined as follows:
\bea \label{rho_hat}
\widehat{\rho}_\bk = \frac{1}{|{\textbf B}_3|} \int_{{\textbf B}_3} \rho(\bz)  \prod_{j = 1}^d  e^{\frac{\;-2\pi i  \;\;k_j}  {b_j-a_j} (z^{(j)} - a_j) } {d}\bz,
\eea
where $|{\textbf B}_3| =\prod_{j = 1}^d (b_j-a_j)$ is the volume.
The above integral is then approximated by a trapezoidal rule, which can help achieve spectral accuracy, and the summation is accelerated by Fast Fourier Transform (FFT) \cite{ST}.

Plugging \eqref{FourSeriB3} into \eqref{I1InteSt3}, we  have 
\bea
I_1(\bx) & = &\sum_{q= 0}^S w_q \int_{{\textbf B}_2} e^{-\tau_q^2 |\by|^2} \rho(\bx-\by) {d} \by\\
& = & \sum_{q = 0}^S w_q  \sum_{\bk} \widehat{\rho}_\bk \prod_{j = 1}^d  e^{\frac{\;2\pi i  \;\;k_j}  {b_j-a_j} (x^{(j)} - a_j)}  \int_{{\textbf B}_2} e^{-\tau_q^2 |\by|^2}  \prod_{j = 1}^d  e^{\frac{-2\pi i  \;\;k_j \; y^{(j)}}  {b_j-a_j} } {d }\by\\
& = & \sum_{q = 0}^S w_q  \sum_{\bk} \widehat{\rho}_\bk \prod_{j = 1}^d  e^{\frac{\;2\pi i  \;\;k_j}  {b_j-a_j} (x^{(j)} - a_j)}  \prod_{j = 1}^d  \int_{-2}^2 e^{-\tau_q^2 |y^{(j)}|^2}   e^{\frac{-2\pi i \,\, k_j \; y^{(j)}}  {b_j-a_j} } {d } y^{(j)}\quad \quad \\
\label{I1FFT}& = & 
 \sum_{\bk}   \widehat{\rho}_\bk \left( \sum_{q= 0}^S w_q    G_\bk^q \right ) \prod_{j = 1}^d  e^{\frac{\;2\pi i  \;\;k_j}  {b_j-a_j} (x^{(j)} - a_j)},\eea
where 
\bea\label{tensor}
G_\bk^q&=& \prod_{j = 1}^d  \int_{-2}^2 e^{-\tau_q^2 |y^{(j)}|^2}\, e^{\frac{-2\pi i  k_j \; y^{(j)}}  {b_j-a_j} } {d } y^{(j)} \nonumber \\
&= &\prod_{j = 1}^d  \int_{0}^2 2\,e^{-\tau_q^2 |y^{(j)}|^2}\, \cos(\tfrac{2\pi  k_j \; y^{(j)}}{b_j-a_j}) {d } y^{(j)}. 
\eea
The coefficients in \eqref{tensor} are tensor products for any fixed index $q$. Notice that $G_\bk^q$ does not depend on the mesh size $\vec{h}:=(h^1,\ldots,h^d)^T$ or the density $\rho$. Therefore,
it can be pre-computed, which greatly enhances the efficiency of the evaluation of $I_1$, because the potential is usually solved many times in simulations. For a given discretization, 
we can pre-compute and store the sums of coefficients, i.e., $\sum_{q= 0}^S w_q    G_\bk^q$, which helps decrease the CPU-time dramatically at a small expense of storage. 
To compute $G_\bk^q$, we only need to calculate three 1-dimensional vectors whose components are given as integrals. 
The integrals in  \eqref{tensor} can be evaluated numerically by a Gauss-Kronrod quadrature up to machine precision \cite{quadpack}. 
Once $\sum_{q= 0}^S w_q    G_\bk^q$ is known, $I_1(\bx)$ can be computed by \eqref{I1FFT} and the summation can be accelerated by FFT. 

\begin{figure}[t!]
\centering\psfig{figure=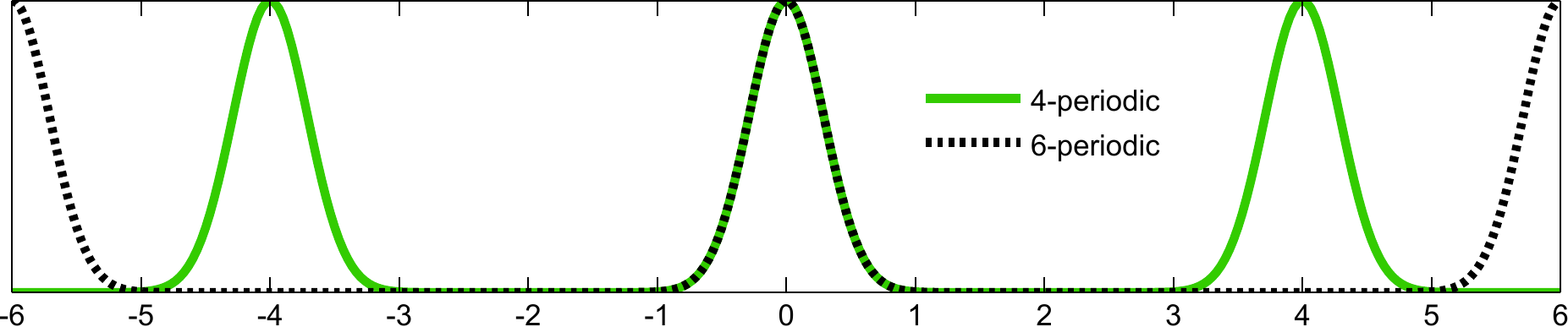,height=2.5cm,width=12cm} 
\caption{The $4$-periodic (green-solid) and the $6$-periodic (black-dotted) extension coincide on $\textbf{B}_3$.}
\label{fig_periodic}
\end{figure}

\begin{remark} Actually, we can restrict the zero-padding to ${\textbf B}_2$ and apply the Fourier series approximation  \eqref{FourSeriB3} on ${\textbf B}_2$ instead of ${\textbf B}_3$.
This can be inferred from the  fact that the $4$-periodic and the $6$-periodic extension of the density coincide on $\textbf{B}_3$, cf. Fig.~\ref{fig_periodic}. Correspondingly, the constants in \eqref{FourSeriB3} will be changed to 
$a_j = -2, b_j =2, \widetilde{n_j} = 2\, n_j, j= 1,\ldots,d$.
\end{remark}

\subsection{Evaluation of the correction integral $I_2(\bx)$}\label{corr}

To evaluate $I_2(\bx)$, we first split it into two integrals as 
\bea
I_2(\bx) &=&\int_{\bR^d}\left( U(\by)- U_{GS}(\by) \right) \rho(\bx-\by) {d} \by, \quad \bx\in \bau  \\
 &=&  \left( \int_{{\mathcal B}_\delta} + \int_{{\mathbb R}^d \backslash{\mathcal B}_\delta} \right) \left( U(\by)- U_{GS}(\by) \right) \rho(\bx-\by) {d} \by, \qquad  \bx\in \bau \\
 \label{splitI2}&:=& I_{2,1}(\bx) + I_{2,2}(\bx), \qquad  \bx\in \bau.  \eea

\vspace{0.05cm}

As can be inferred from the compactness assumption, i.e., $\textrm {supp}(\rho)\subset {\textbf B}_1$, 
we have for $\bx \in \textbf{B}_1$ that $\textrm {supp} \{\rho(\bx-\by)\} \subset {\textbf B}_2$. Therefore,
the latter integral $I_{2,2}(\bx)$ is equivalent to an integral defined on a bounded domain, i.e., ${\mathcal D}:={\textbf B}_2\backslash {\mathcal B}_\delta$.
To be more precise, we have 
\bea
I_{2,2}(\bx) &=& \int_{ {\mathbb R}^d \backslash{\mathcal B}_\delta }\left( U(\by)- U_{GS}(\by)  \right) \rho(\bx-\by) {d} \by , \quad \bx \in {\textbf B}_1\\
&=&  \int_{{\mathcal D} }\left(U(\by)- U_{GS}(\by)   \right) \rho(\bx-\by) {d} \by , \qquad \bx \in {\textbf B}_1.
\eea
Since the GS approximation of $U$ gives an error $\varepsilon$ on $\mathcal{D}$, by adopting in spherical/polar coordinates, we have 
\bea\label{I_22est}
|I_{2,2}(\bx) | & \leq& |S^{d-1}| \, \left\| \rho \right\|_\infty \,\int_\delta^2 r^{d-1}\,\left| U(r) - U_{GS}(r)\right| {d}r\\
&\leq & C \left\| \rho \right\|_\infty \max_{r \in [\delta,2]} |\left( U(r) - U_{GS}(r) \right)|, 
\eea
where $|S^{d-1}| = \frac{2\pi^{d/2}}{\Gamma(d/2)}$ is the volume of unit surface in $\mathbb R^d$ and  $C$ is a constant not depending on the density. Then we neglect $I_{2,2}$ because of the near-machine precision accurate GS approximation  in \eqref{I_22est}. We refer to Sec.~\ref{kernel} and Fig.~\ref{fig_Gaussian_Approx} for more details.

\vspace{0.5cm}
\begin{remark}
In \eqref{I_22est}, the error estimate does not really have to depend on the density $\rho$, simply because 
we can normalize the density to be $\|\rho\|_\infty =1$.
\end{remark}
\vspace{0.5cm}

 In order to compute $I_{2,1}$, we need first to interpolate the density function in a $\delta$-neighborhood of $\bx$. Since $\delta$ is small (we choose $\delta = 10^{-3} \text{ or }10^{-4}$ in our implementation), 
the interpolation of the density $\rho_\bx(\by) :=\rho(\bx-\by)$ within ${\mathcal B}_\delta$ can be done by the Taylor expansion. To be exact, we have 
\bea
\rho_{\bx}(\by)=  \mathrm P_\bx(\by)  + \mathrm R_\bx(\by),\quad  \by \in {\mathcal B}_\delta,
\eea
where $\mathrm P_\bx(\by)$,  the third order Taylor expansion, is defined as follows:
\bea\label{taylor} 
 \mathrm P_\bx(\by)&=& \rho_{\bx}(\textbf{0}) + \sum_{j=1}^d \frac{\partial \rho_{\bx}(\textbf{0})}{\partial y_j} y_j
+ \frac{1}{2}\sum_{j,k=1}^d \frac{\partial^2 \rho_{\bx}(\textbf{0})}{\partial y_j \partial y_k} y_j\, y_k \nonumber \\
 &&+
\frac{1}{6}\sum_{j,k,l=1}^d \frac{\partial^3 \rho_{\bx}(\textbf{0})}{\partial y_j \partial y_k \partial y_l} y_j\, y_k\, y_l,\eea
and the remainder $\mathrm R_\bx(\by) = C(\rho,\bx) |\by|^4$ with the constant $C(\rho,\bx)$ depending on the 
density $\rho$ and $\bx$.
Spherical coordinates are now used in order to remove the singularity. 
Next, we plug the spherical representation of \eqref{taylor} into \eqref{splitI2}. 
After integration over the $r, \theta, \phi$ respectively, the evaluation of $I_{2,1}$ comes down to 
simple multiplication of $\Delta \rho$ and some constants, since the contributions of the odd derivatives in \eqref{taylor} and off-diagonal components of the Hessian vanish. It is noteworthy to point out that we do not have to resort to any numerical quadrature here.

The approximation error of $I_{2,1}$ by $\widetilde{I_{2,1}}$ (the integral with Taylor expansion) is estimated as follows:
\bea
|(I_{2,1} -\widetilde{I_{2,1}} )(\bx) |&= &\Big| \int_{{\mathcal B}_\delta} \left( U(\by) - U_{GS}(\by) \right) C(\rho,\bx) |\by|^4  {d} \by\Big|\\
&\leq & \|C(\rho,\bx)\|_\infty |S^{d-1}|\,\Big|  \int_0^\delta  r^{d-1} \, r^4 | U(r)-  U_{GS}(r) |  {d} r  \Big| \\
&\leq&  \|C(\rho,\bx)\|_\infty \,  |S^{d-1}| \, C_S\, 
\left\{\begin{array}{c c} 
\delta^{d+3}, & \text{Coulomb kernel} \\
\delta^6 \,|\log\delta|, & \text{Poisson kernel}   
\end{array}\right.
\eea
where 
\bea\label{I_2h}
\widetilde{I_{2,1}}(\bx)  =  \int_{{\mathcal B}_\delta} \left( U(\by)- U_{GS}(\by) \right) \mathrm P_\bx(\by) {d} \by,
\eea
and  $C_S$ is a positive parameter depending on the weights of the Gaussian-sum approximation (cf. Sec.~\ref{kernel}).
\vspace{0.5cm}
\begin{remark}
The main computational work relies on the FFT. Since there are several successful versions of distributed-memory parallel FFT implementations, e.g., parallel version of the FFTW and its extensions \cite{fftw97,pfft}, the performance 
of our method can be enhanced with such libraries.
\end{remark}

\subsection{Extension to the dipolar potential}\label{dipole}
The dipolar potential is of great importance in condensed matter and quantum mechanics \cite{YY2000,YY2001}. It also takes convolution form, 
i.e., $u(\bx) = U \ast \rho$ where  
\begin{eqnarray}\label{dipkerneld}
\qquad \quad  U(\bx) &=& \frac{3}{4\pi}\frac{\bm\cdot\bn-3(\bx\cdot \bn)
(\bm\cdot\bx)/|\bx|^2}{|\bx|^3} \\
\nonumber
&=& -(\bm \cdot \bn )\delta(\bx)-3\, \partial_{\bn\bm}\left(\frac{1}{4\pi|\bx|}\right),\ \bx \in \bR^3.
\end{eqnarray}
Using the convolution theorem, we can rewrite the dipolar potential as follows:
\bea
u(\bx) &=& -(\bm \cdot \bn ) \rho(\bx) + \partial_{\bn\bm}\left(\frac{1}{4\pi|\bx|} \right) \ast \rho=-(\bm \cdot \bn ) \rho(\bx) + \frac{1}{4\pi|\bx|} \ast \left(\partial_{\bn\bm}\rho \right).\quad \quad 
\eea
Therefore, the computation of $u$ consists of the evaluation of the 3D Coulomb potential with the source term $\partial_{\bn\bm}\rho(\bx) $. Since the density $\rho(\bx)$ is smooth and compactly supported
in ${\textbf  B}_L$, it can be approximated by finite Fourier series with spectral accuracy, and so does the second derivative $\partial_{\bn\bm}\rho(\bx)$.
The source term $\partial_{\bn\bm}\rho(\bx) $ can be easily computed with arithmetic operations of the discrete Fourier coefficients.

We note that a similar situation arises in the \textit{Davey-Stewartson} nonlocal potential in $2$D, 
where one could solve Poisson's equation with the second derivative of the density as source term. Hence, in the convolution form, 
one only has to convolve the 2D Poisson kernel with the second order derivative of the density, cf. example~\ref{ds} in Sec.~\ref{numerics}.

\vspace{0.5cm}

For readers' convenience, we summarize the key steps of our algorithm in Algorithm \ref{method_alg}.
\begin{algorithm}
\caption{Evaluation of the nonlocal potential \eqref{ConvGen}}
\label{method_alg}
 \textbf{Precomputation}
\begin{enumerate}
\item{ Gaussian-sum approximation of the kernel $U(x)$ in \eqref{kernelPhy}.}
\item{ Fourier coefficients $G_\bk^q$ in \eqref{tensor} via its tensor product composing vectors.}\end{enumerate}
\textbf{Actual computation}
\begin{enumerate}
\item{Compute $\widehat{\rho}_\bk$, cf. \eqref{rho_hat}.}
\item{Evaluate $I_1$ by \eqref{I1FFT} with FFT.}
\item{Compute the Laplacian of $\rho$ with FFT.} 
\item{Evaluate $\widetilde{I_{2,1}}$ by \eqref{I_2h}.}
\item{Add $I_1$ and $\widetilde{I_{2,1}}$  to obtain the approximation of $u$.}
\end{enumerate}
\end{algorithm}
\subsection{Anisotropic densities}
For clarity, we assume that the (rescaled) density is compactly supported in the rectangular box $\textbf{B}_{1,\eta} := [-1,1]^{d-1} \times \eta [-1,1]$. Here, the regular integral and correction integral in Sec.~\ref{I1} and \ref{corr} needs modifications accordingly.
More precisely, for regular integral evaluation, cf. Sec.~\ref{I1}, the related changes are listed as follows:
\bea
I_1(\bx) = \sum_{\bk}   \widehat{\rho}_\bk \left( \sum_{q = 0}^S w_q    G_\bk^q \right ) e^{\frac{\;2\pi i  \;\;k_d}  {\varepsilon(b_d-a_d)} (x^{(d)} - \varepsilon a_d)}\,\prod_{j = 1}^{d-1}  e^{\frac{\;2\pi i  \;\;k_j}  {b_j-a_j} (x^{(j)} - a_j)},
\eea
where
\bea\label{tensor_ani}
G_\bk^q&=& \int_{-2\eta}^{2\eta} e^{-\tau_q^2 |y^{(d)}|^2}   e^{\frac{-2\pi i  k_d \; y^{(d)}}  {\eta (b_d-a_d)} } {d } y^{(d)} \, \prod_{j = 1}^{d-1}  \int_{-2}^{2} e^{-\tau_q^2 |y^{(j)}|^2}   e^{\frac{-2\pi i  k_j \; y^{(j)}}  {b_j-a_j} } {d } y^{(j)}.
\eea
For the correction integral $I_2$, one has to choose $\delta$ smaller than $\eta$ so as to guarantee the validity and accuracy in the Taylor expansion. 
Numerical results for the 2D/3D Coulomb potentials are displayed in Section \ref{numerics}, cf. Tab.~\ref{Tab_Coulomb2D_Aniso} and Tab.~\ref{Tab_Coulomb3D_Aniso}.

\begin{remark}
Given a general rectangular domain, e.g., the 2D $[-L_x,L_x]\times [-L_y,L_y] $, the above algorithm adapted for anisotropic densities can then be adapted simply by setting $\eta = \min\{L_x/L_y,L_y/L_x\}$. The 3D Coulomb potential evaluated on a more general rectangular domain 
can also be adapted similarly.
\end{remark}

\section{Kernel approximation}\label{kernel}
The kernel's high-resolution approximation is of great importance in our algorithm. Here we choose the effective 
GS approximation, which has already been exploited extensively in \cite{KernelGauss1,KernelGauss,Braess_expsum,steger_1993}. 
Its tensor product structure leads to a considerable simplification of the pre-computation \eqref{tensor} in the regular integral $I_1$ evaluation. 
The higher resolution in the GS approximation achieved with sinc quadrature allows us to neglect the integral $I_{2,2}$, cf. \eqref{I_22est}, thus, confines the near-field correction integral into a small ball $\mathcal B_\delta$.

The sinc quadrature approach to obtain the GS approximation relies on a Gaussian integral representation 
of the kernel $U$. In this section, we briefly review some facts of the \textit{sinc-quadrature} \cite{HackBush2005,steger_1993} to make our paper reasonably self-contained, 
then present the concrete approximations of the kernels $1/r$ and $\ln r$ on an interval $[\delta, 2],\, 0<\delta\ll 1$.  
\subsection{Sinc quadrature}\label{sincquad} The sinc function $\text{sinc}(t) := \frac{\sin(\pi t)}{\pi t}$ is an analytic function, which equals to $1$ at $t=0$ 
and zero at $t \in \mathbb{Z}\setminus\{0\}$. Sufficiently fast decaying continuous functions $f\in C(\mathbb{R})$ can be interpolated at the grid points 
$t_k = k\vartheta\in \vartheta\mathbb{Z},\, \vartheta>0$ (step size) by functions $\mathcal{S}_{k,\vartheta}(t) := \text{sinc}(t/\vartheta - k)$, i.e.,
\bea
f_\vartheta(t) = \sum_{k\in \mathbb{Z}} f(k\vartheta)\mathcal{S}_{k,\vartheta}(t).
\eea
Since $\int_{\mathbb{R}} \text{sinc}(t)\,dt = 1$, an interpolatory quadrature for $\int_{\mathbb{R}} f(t) \, dt$ is given as follows:
\begin{align}\label{sinc_int}
\int_{\mathbb{R}} f(t) \, d t \approx \vartheta \sum_{k\in \mathbb{Z}} f(k\vartheta),
\end{align}
which can be viewed as ``infinite trapezoidal rule'' quadrature. Finite truncation to the first $2S+1$ terms, i.e., $k = -S, \ldots,S$, of the infinite sum leads to the \textit{sinc quadrature rule} with the error $\vartheta \sum_{|k|>S} f(k\vartheta)$ depending on the decay-rate of $f$. 
For functions $f(z)$ in the Hardy space $H^1(D_\lambda),\, \lambda < \pi/2$, that is to say,  $f(z)$ is holomorphic in the strip $D_\lambda := \{z\in \mathbb{C}:\,|\Im\,z| \leq \lambda\}$ and  
\begin{align}
N(f,D_\lambda) := \int_{\partial D_{\lambda}} |f(z)| \, |dz| = \int_{\mathbb{R}} \big(|f(t + i\lambda)| + |f(t - i\lambda)| \big)\, \text{d}\,t < \infty,
\end{align} 
and if $f(z)$ also satisfies the double exponential decay property on the real axis, we have the following exponential error estimate for sinc quadrature approximation, see \cite{HackBush2005} (Proposition 2.1).

\vspace{0.5cm}

\begin{prop}[\cite{HackBush2005}]\label{sinc_theo}
Let $f \in H^1(D_\lambda)$ with $\lambda < \pi/2$. If $f$ satisfies the double exponential decay condition, i.e.,
\begin{align}\label{double_exp}
|f(t)| \leq C\,\exp(-be^{a|t|}) \quad \forall \; t\in \mathbb{R} \,\, \text{ with }\,\, a,b,C > 0,
\end{align}
then the quadrature error for the special choice $\vartheta = \ln(\frac{2\pi a S}{b})/(aS)$ satisfies
\begin{align}\label{quad_err}
\left|\int_{\mathbb{R}} f(t) \, d t - \vartheta \sum_{|k|\leq S} f(k\vartheta)\right| \leq C\, N(f,D_\lambda)\, \exp\Big(\frac{-2\pi \lambda a S}{\ln(2\pi aS/b)}\Big).
\end{align}
\end{prop}
\begin{remark}
In the case of an integral expression $\int_{\mathbb{R}} g(t)\,e^{x h(t)} \,dt$,  the constants in \eqref{quad_err} depend on the parameter $x$. 
For some fixed $x$, an accuracy of $\varepsilon > 0$ can be achieved with $S:= \mathcal{O}(|\ln\varepsilon|\cdot \ln|\ln\varepsilon|)$. 
Moreover, in our computations, we use the simplified step-size $\vartheta = c_0 \ln(S)/S$, cf. \eqref{quad_err}, as in \cite{HackBush2005} with some 
positive constant $c_0$ (i.e., $c_0 = 2.1$ for the Coulomb kernel and $c_0 = 1$ for the Poisson kernel).
\end{remark}
\vspace{0.25cm} 

In the following, we first represent the kernels $1/|\bx|$ and $\ln |\bx|$ in Gaussian integral form and then apply the sinc-quadrature to obtain a GS approximation. 
These approximations are valid in an interval $[\delta,2]$ and used to split the convolution \eqref{ConvGen} into a regular integral and short-range correction integral, cf. \eqref{key_form}.

\subsection{Approximation of the Coulomb kernel $1/r$ over $[\delta,2]$}\label{1_r}
Starting from the following identity
\begin{align}\label{Gauss_int}
 \int_0^\infty \tau^\alpha e^{-\rho \tau^2}\,{d}\tau = \Gamma (\tfrac{\alpha+1}{2}) \rho^{-\tfrac{\alpha+1}{2}},\,\, \rho >0,\, \alpha>-1,
\end{align}
for $\alpha = 0$ and $\rho = |\mathbf{x}|^2$, we have a Gaussian integral representation
\begin{align}\label{GauInte}
 \frac{1}{|\mathbf{x}|} = \frac{2}{\sqrt\pi} \int_0^\infty e^{-|\mathbf{x}|^2 \tau^2}\,{d} \tau = \frac{2}{\sqrt\pi} \int_0^\infty \prod_{p = 1}^d  e^{-{x^{(p)}}^2 \tau^2}\,{d} \tau.
\end{align}
Applying some numerical quadrature to the integral $\int_{0}^{\infty} e^{\rho \tau^2} {d} \tau$ leads to a GS approximation
\begin{align}\label{sinc_quad}
 \frac{1}{|\mathbf{x}|} \approx \sum_q w_q \prod_{j = 1}^d  e^{-\tau_q^2\, {x^{(j)}}^2 }.
\end{align}
\begin{remark}
For kernels $1/|\mathbf{x}|^{\beta},\,\beta > 0$, choosing $\alpha = \beta-1$, formula \eqref{Gauss_int} gives
a Gaussian integral representation similar to that of the Coulomb kernel. Substituting  $\rho=|\mathbf{x}|^2$  and applying the numerical quadrature, we obtain a  similar GS approximation.
\end{remark}
\vspace{0.5cm}

The numerical quadrature we choose here is the \textit{sinc quadrature}, cf. Sec.~\ref{sincquad}, which is suited for integrals $\int_{\mathbb{R}} f(t) \, {d} t$ with $f \in C(\mathbb{R})$ decaying sufficiently fast. 
More precisely, by a change of variables in \eqref{GauInte}, i.e., $\tau = \sinh t:=\frac{1}{2}(e^t -e^{-t})$, 
the updated integrand, now a function in Hardy space with double exponential decay on the real axis, 
satisfies the condition of Proposition~\ref{sinc_theo}. A sinc quadrature applies readily and  the quadrature converges exponentially with respect to the number of Gaussian terms.
The integrand is an even function, and we shall end up with only $S+1$ terms.

\begin{figure}[t!]
\centerline{ \psfig{figure=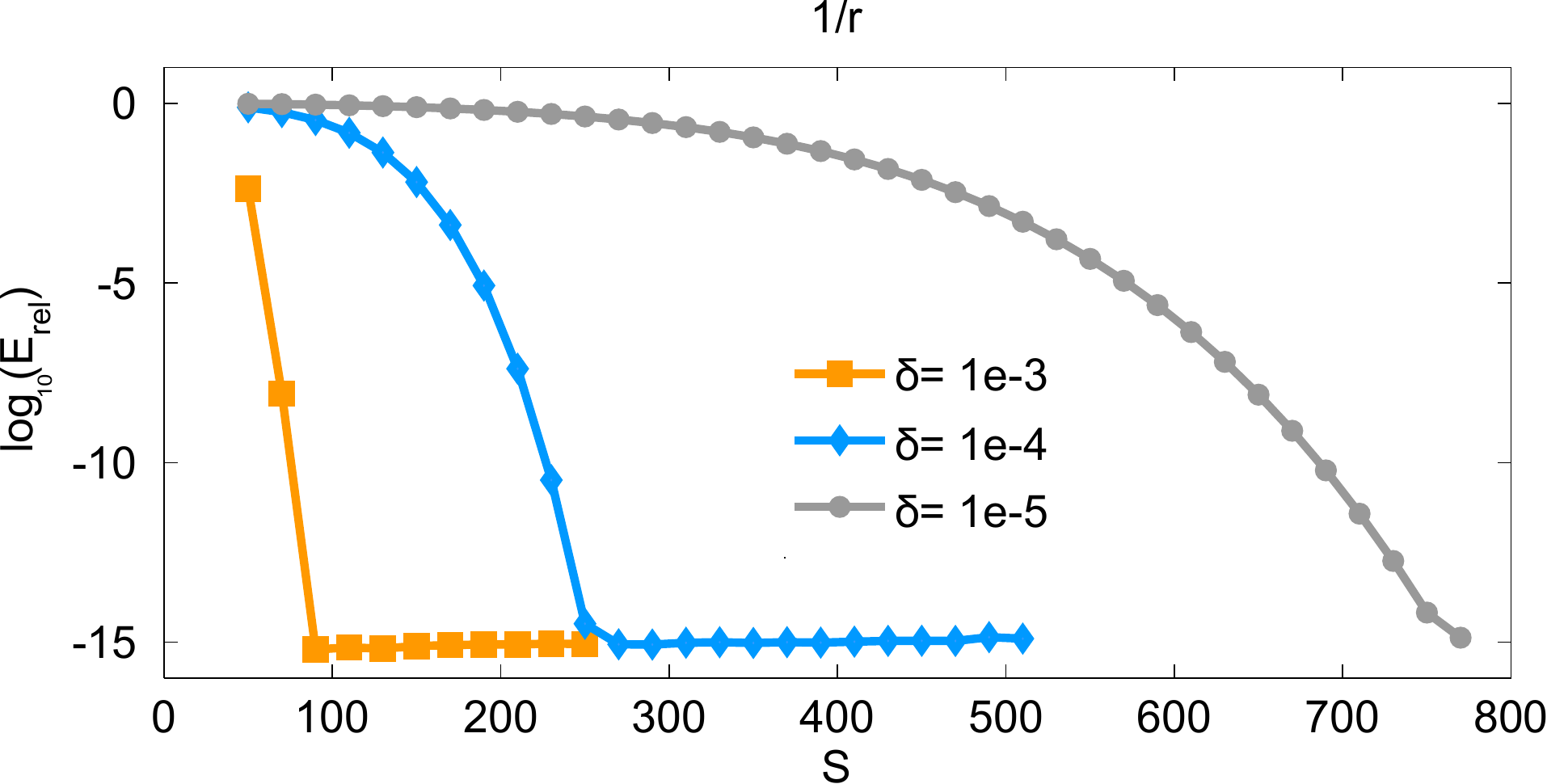,height=4.0cm,width=6.5cm}
\psfig{figure=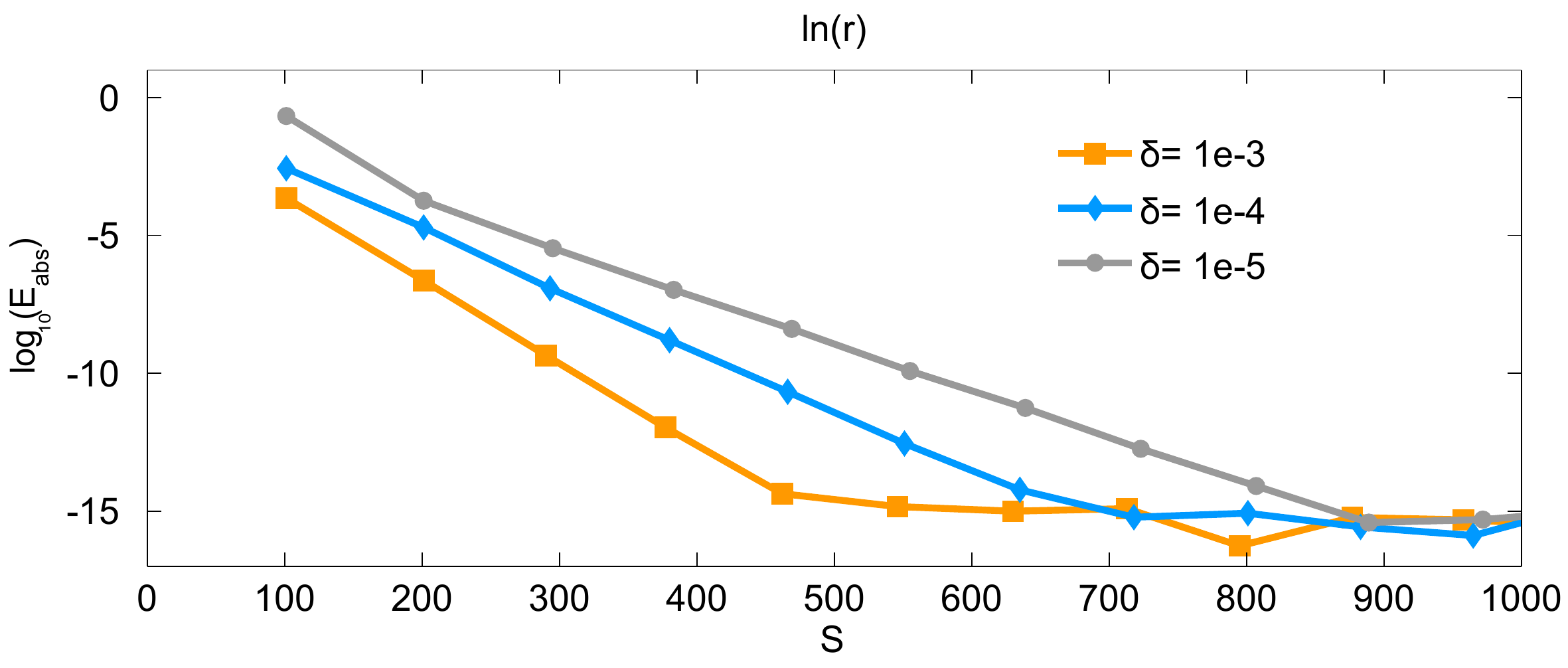,height=4.0cm,width=6.8cm} }
\caption{Number of terms $S$ versus $E_{\rm rel}$ for the kernel $1/r$ (left), $E_{\rm abs}$ for $\ln r$ (right) on $[\delta,2]$.}
\label{fig_Gaussian_Approx}
\end{figure}

A detailed analysis (similar to that in \cite{exl_2014}) shows that the quadrature for $1/r, r = |\bx|$ is acceptable for an interval $r \in [\delta , 2], \, 0 < \delta \ll 1$.
The left picture of Fig.~\ref{fig_Gaussian_Approx} shows the relative error $E_{\rm rel}$  of the GS approximation, where  $E_{\rm rel} :=\|\,1- \sum_{q= 0}^S w_q\, r \,e^{-\tau_q^2 r^2}\|_{L^\infty( (\delta,2] ) }$, 
from where one could observe the high-resolution approximation.

\subsection{Approximation of the Poisson kernel $\ln r $ over $[\delta,2]$}\label{logr}
In the subsection, we shall present a GS approximation for the 2D Poisson kernel $\ln|\bx| := \ln{\sqrt{{x^{(1)}}^2 + {x^{(2)}}^2}}$.  Setting $\alpha = 1$ in \eqref{Gauss_int}, we have 
\begin{align}\label{1r_int}
\frac{1}{{x^{(1)}}^2 + {x^{(2)}}^2} = \int_0^\infty e^{-({x^{(1)}}^2 + {x^{(2)}}^2) \tau}\,{d} \tau.
\end{align}
Applying a change of variables $\tau = \ln{(1 + \exp(\sinh t))}$, the integration domain  in \eqref{1r_int} is now the whole real axis and the integrand has double exponential decay. 
Thus, we can apply the sinc-quadrature ($2S+1$ terms in this case)  to obtain a GS approximation of $|\bx|^{-2}$ in $[1,R],\, R>1$, from which we can change to the interval $[\delta,2]$ following a scaling argument.
Inserting the GS approximation of $1/|\bx|^2$ into the following formula
\begin{align}
\ln{ \sqrt{{x^{(1)}}^2 + {x^{(2)}}^2}} = \int_{\sqrt{1-{x^{(2)}}^2}}^{x^{(1)}} \frac{y}{y^2 + {x^{(2)}}^2} \,{d}y,
\end{align}  
we obtain an GS approximation as follows:
\begin{align}\label{log_approx}
\ln{ \sqrt{{x^{(1)}}^2 + {x^{(2)}}^2}} \approx C_0  -  \sum_{q=1}^S \widetilde{w}_q \, e^{-\widetilde{\tau}_q ({x^{(1)}}^2 + {x^{(2)}}^2)} =: \sum_{q=0}^S w_q \, e^{-\tau_q^2 ({x^{(1)}}^2 + {x^{(2)}}^2)},
\end{align}
where $w_0 = C_0$, $w_q = -\widetilde{w}_q,\, q\geq 1$ and $\tau_0 = 0,\, \tau_q = {\widetilde{\tau}_q}^{1/2},\,q\geq 1$. 
We point out that the coefficients $w_q$ and $\tau_q$ in \eqref{log_approx} should be computed stably (double precision) for both small and large nodes $\tau_q$.
The right figure in Fig.~\ref{fig_Gaussian_Approx} shows the absolute error $E_{\rm abs} :=\|\,\ln r- \sum_{q=0}^S w_q\,e^{-\tau_q^2 r^2}\|_{L^\infty( (\delta,2] ) }$ for the kernel $\ln r$ on $[\delta,2]$.

 \section{Numerical results}\label{numerics}
In order to demonstrate the accuracy and efficiency performance of our method, we perform several numerical tests in this section. All the numerical errors are calculated in the relative maximum norm, which is defined as follows
\be
\label{error_infty}
E:=\frac{\|u-u_{\vec h}\|_{l^\infty}}{\|u\|_{l^\infty}}=
\frac{\max_{\bx\in \mathcal{T}_h} |u(\bx)-u_{\vec h}(\bx)|}{\max_{\bx\in \mathcal{T}_h} |u(\bx)|},
\ee
where $\mathcal{T}_h$ is the rectangular computational domain discretized uniformly in each direction with mesh sizes ${\vec h}=(h_x,h_y)^T \text{ and } (h_x,h_y,h_z)^T$ for 2D and 3D, respectively.
The grid function $u_{\vec h}$ is the numerical solution and $u$ is the exact solution. 
Further, we denote the total number of grid points by $N := n_x n_y n_z$ and $N := n_x n_y$ for the 3D and 2D domain, respectively. 
For the sake of convenience, the mesh size in the $p$-th direction $h_p$ is simply denoted by $h$ hereafter.\\

The algorithm is implemented in Fortran, the code is compiled by ifort (version 14.0.2) using the option -O3, and executed on 64-bit Linux on a 2.53GHz Intel(R) Xeon(R) E5540 CPU with 6MB cache. 
The CPU times shown in this section do not include all the pre-computation times. The exclusion of all the pre-computation steps is justified by the fact that in many applications/real simulations, one  needs to evaluate the nonlocal potential multiple times on the same grid.

\vspace{0.5cm}
\begin{exmp}\label{Coul3D} The 3D Coulomb potential. \end{exmp} 
For the density $\rho(\bx):= e^{-(x^{2}+y^2+\gamma^2z^2)/\sigma^2}$ with $\sigma>0$ and $\gamma\ge 1$, the
3D Coulomb potential, with the kernel $U(\bx) = \frac{1}{4\pi |\bx|}$, can be computed analytically as
\be\label{DensGaus}
u(\bx) =  \left\{\ba{ll}
\frac{\sigma^3 \sqrt{\pi}}{4\;|\bx|\;}\,
\text{Erf}\left( \frac{|\bx|}{ \sigma }\right), &\gamma = 1,\\[1em]
\frac{\sigma^2}{4\gamma} \int_0^\infty \frac{e^{-\frac{x^2+y^2}{\sigma^2(t+1)}}
e^{-\frac{z^2}{\sigma^2(t+\gamma^{-2})}}} {(t+1)\sqrt{t+\gamma^{-2}}}dt     , & \gamma  \ne 1,
\ea\right. \qquad \bx\in{\mathbb R}^3,
\ee
where $ \textrm{Erf}(x) =\frac{2}{\sqrt{\pi}} \int_0^x e^{-t^2} d t $ for $x\in \mathbb R$ 
is the error function. For densities $\rho_{\bx_0 }(\bx) := \rho(\bx -\bx_0)$ with $\bx_0 \in \mathbb R^3$, the corresponding 
3D Coulomb potential is given exactly as $u_{\bx_0}(\bx) = u(\bx-\bx_0)$. 

The 3D Coulomb potential is computed on $[-L,L]^2\times [-L/\gamma, L/\gamma]$ with mesh size $ h_x = h_y, h_z = h_x/\gamma$. 
Table \ref{Tab_Coulomb3D} shows the error $\erh$ and computation time for the isotropic density, i.e., $\gamma = 1$, with $\sigma = 1.2$ on different domains $[-L,L]^3$, where $T_1,T_2$ and $T_{total}$ denote 
hereafter the time for the evaluation of $I_1, I_2$ in \eqref{key_form} and the total time, respectively. Table \ref{Tab_Coulomb3D_shift} presents the results of the potential for shifted density 
with $\sigma = 1.2$ and $\bx_0=(1,2,1)^T$ computed on $ [-12,12]^3$.
Table \ref{Tab_Coulomb3D_Aniso} lists the errors $\erh$ and timings for different anisotropic densities with $\sigma = 2$ computed on $[-12,12]^2\times  [-12/\gamma,12/\gamma] $ using the same mesh size in x and y-direction, 
i.e., $h_x = h_y= 1/4$ and a different mesh size in z-direction, i.e., $h_z = h_x/\gamma$.

From Tab.~\ref{Tab_Coulomb3D}-\ref{Tab_Coulomb3D_Aniso}, we can conclude that:
(i) The method is spectrally accurate with respect to the mesh size $h$ and efficient with a complexity of 
$O(N\ln N)$;  (ii) The anisotropic potential can be computed with spectral accuracy without increasing the memory or CPU time as the $\gamma$ tends larger, thus, it is ideal for applications.

\begin{table}[h!]
\tabcolsep 0pt \caption{Errors and timings of the 3D Coulomb potential in Example \ref{Coul3D} with  isotropic density with $\sigma = 1.2$ on $[-L,L]^3$. } \label{Tab_Coulomb3D}
\begin{center}\vspace{-1em}
\def\temptablewidth{1\textwidth}
{\rule{\temptablewidth}{1pt}}
\begin{tabularx}{\temptablewidth}{@{\extracolsep{\fill}}p{1.05cm}clllll}
$L =8$&$N$&  $\erh$ & $T_1$ & $T_2$ & $T_{total}$ \\ \hline
$h = 1$         & $16^3$  &1.096E-03 &9.99E-04 &1.00E-03&2.00E-03\\
$h\!\! =\!\!1/2$& $32^3$  &1.130E-09 &1.60E-02 &2.00E-03&1.80E-02\\
$h\!\! =\!\!1/4 $& $64^3$ &6.169E-16 &1.93E-01 &1.90E-02&2.12E-01\\
$h \!\! =\!\!1/8$& $128^3$&6.187E-16 &1.69     &6.28E-01&2.31\\
$h \!\! =\!\!1\!/\!16$& $256^3$&7.725E-16  &15.03  &4.71&19.74\\
\hline \hline
$L\!=\!16$&$N$&  $\erh$ & $T_1$ & $T_2$ & $T_{total}$ \\ \hline
$h = 1$         & $32^3$  & 1.113E-03 &1.60E-02 &2.00E-03& 1.80E-02 \\
$h\!\! =\!\!1/2$& $64^3$  & 1.191E-09 &1.95E-01 &2.10E-02& 2.16E-01 \\
$h\!\! =\!\!1/4 $&$128^3$ & 9.259E-16 &1.71     &6.22E-01& 2.33     \\
$h \!\! =\!\!1/8$&$256^3$ & 9.271E-16 &15.18    &4.76    & 19.94    \\
\end{tabularx}
{\rule{\temptablewidth}{1pt}}
\end{center}
\end{table}

\begin{table}[h!]
\tabcolsep 0pt \caption{Errors and timings of the 3D Coulomb potential in Example \ref{Coul3D} for shifted Gaussian density with $\sigma = 1.2$ and $\bx_0=(1,2,1)^T$
on $[-12,12]^3$. } \label{Tab_Coulomb3D_shift}
\begin{center}\vspace{-1em}
\def\temptablewidth{1\textwidth}
{\rule{\temptablewidth}{1pt}}
\begin{tabularx}{\temptablewidth}{@{\extracolsep{\fill}}p{1.05cm}clllll}
$L \!\!=\!12$&$N$&  $\erh$ & $T_1$ & $T_2$ & $T_{total}$ \\ \hline
$h = 1$          &      $24^3 $ & 1.108E-03  & 7.00E-03  & 4.00E-03  & 1.10E-02\\
$h\!\! =\!\!1/2$ &      $48^3 $ & 1.175E-09  & 8.10E-02  & 1.20E-02  & 9.30E-02\\
$h\!\! =\!\!1/4 $&      $96^3 $ & 6.182E-16  & 7.03E-01  & 1.08E-01  & 8.11E-01\\
$h \!\! =\!\!1/8$&      $192^3$ & 7.717E-16  & 6.30      & 1.08      & 7.37 \\
\end{tabularx}
{\rule{\temptablewidth}{1pt}}
\end{center}
\end{table}

\begin{table}[h!]
\tabcolsep 0pt \caption{Errors and timings of the 3D Coulomb potential in Example \ref{Coul3D} for anisotropic densities with $\sigma = 2$ computed on $ [-12,12]^2\times \frac{1}{\gamma} [-12,12] $ with $h_x =h_y =1/4, h_z = h_x/\gamma$ ($N = 96^3$). } \label{Tab_Coulomb3D_Aniso}
\begin{center}\vspace{-1em}
\def\temptablewidth{1\textwidth}
{\rule{\temptablewidth}{1pt}}
\begin{tabularx}{\temptablewidth}{@{\extracolsep{\fill}}p{1.05cm}cllll}
$\gamma$ & $\erh$ & $\|u\|_{\rm max}$ &$T_1$ &$T_2$ &$T_{total}$ \\ \hline  
$1$ &4.486E-16 &2     & 6.76E-01 &1.01E-01&7.77E-01\\
$2$ &5.599E-16 &1.209 & 6.83E-01 &1.02E-01&7.85E-01\\
$4$ &1.427E-15 &0.681 & 6.81E-01 &1.00E-01&7.81E-01\\
$8$ &2.606E-14 &0.364 & 6.78E-01 &1.03E-01&7.81E-01\\
\end{tabularx}
{\rule{\temptablewidth}{1pt}}
\end{center}
\end{table}

\vspace{0.5cm}
\begin{exmp} \label{Coul2D} The 2D Coulomb potential. \end{exmp} 
For the density $\rho(\bx)= e^{-(x^{2}+\gamma^2y^2)/\sigma^2}$ with $\sigma>0$ and $\gamma\ge1$,  the 2D Coulomb potential, with the  kernel $U(\bx) = \frac{1}{2\pi |\bx|}$, can be obtained analytically as
\be\label{2.5D-clb-exact}
u(\bx) =  \left\{\ba{ll} \frac{\sqrt{\pi}\, \sigma}{2}
\,{\mathrm I}_0\left(\frac{|\bx|^2}{2 \sigma^2}\right)\,e^{-\frac{|\bx|^2}{2\sigma^2}}, &\gamma = 1,\\[0.45em]
\frac{\sigma}{\gamma \sqrt{\pi}} \int_0^\infty \frac{e^{-\frac{x^2}{\sigma^2(t^2+1)}}
e^{-\frac{y^2}{\sigma^2(t^2+\gamma^{-2})}}} {\sqrt{t^2+1}\sqrt{t^2+\gamma^{-2}}}dt     , & \gamma  \ne 1,
\ea\right.\qquad \bx\in{\mathbb R}^2,
\ee
where ${\mathrm I}_0$ is the modified Bessel function of the first kind \cite{Handbook}.

Similarly, we shall first present the accuracy and efficiency performance of our method on fixed domains $[-L,L]^2$ with $\sigma  = 1.2$ in Table \ref{Tab_Coulomb2D}. Then we compute the 2D Coulomb potential for anisotropic densities on $[-L,L]\times [-L/\gamma, L/\gamma]$  using a fixed mesh size in x-direction, i.e., $h_x  = 1/4$ and $h_y = h_x/\gamma$, 
cf. Table \ref{Tab_Coulomb2D_Aniso}. 
From Tab.\ref{Tab_Coulomb2D} and Tab.\ref{Tab_Coulomb2D_Aniso}, we can draw similar conclusion as that in the 3D Coulomb example and we omit it for brevity. 

\begin{table}[h!]
\tabcolsep 0pt \caption{Errors and timings of the 2D Coulomb potential in Example \ref{Coul2D} for $\sigma = 1.2$ on $ [-L,L] ^2$ with different mesh size.} \label{Tab_Coulomb2D}
\begin{center}\vspace{-1em}
\def\temptablewidth{1\textwidth}
{\rule{\temptablewidth}{1pt}}
\begin{tabularx}{\temptablewidth}{@{\extracolsep{\fill}}p{1.05cm}lllll}
$L =8$&$N$&  $\erh$ & $T_1$ & $T_2$ & $T_{total}$ \\ \hline
$h = 1$         & $16^2$       & 9.426E-04 &0        &0        &0         \\
$h\!\! =\!\!1/2$& $32^2$       & 1.720E-09 &0        &0        &0         \\
$h\!\! =\!\!1/4 $& $64^2$      & 4.190E-16 &2.00E-03 &1.01E-03 &3.00E-03  \\
$h \!\! =\!\!1/8$& $128^2$     & 5.229E-16 &6.00E-03 &2.00E-03 &8.00E-03  \\
$h \!\! =\!\!1\!/\!16$& $256^2$& 5.229E-16 &2.30E-02 &7.01E-03 &3.00E-02  \\
\hline \hline
$L\!=\!16$&$N$&  $\erh$ & $T_1$ & $T_2$ & $T_{total}$ \\ \hline
$h = 1$         & $32^2$       &9.576E-04 &1.00E-03 &0     &1.00E-03  \\
$h\!\! =\!\!1/2$& $64^2$       &1.815E-09 &1.00E-03 &0     &1.00E-03  \\
$h\!\! =\!\!1/4 $&$128^2$      &5.846E-15 &5.00E-03 &2.00E-03 &7.00E-03  \\
$h \!\! =\!\!1/8$&$256^2$      &5.846E-15 &2.60E-02 &7.00E-03 &3.30E-02  \\
$h \!\! =\!\!1\!/\!16$& $512^2$&6.055E-15 &2.47E-01 &2.80E-02 &2.75E-01  \\
\end{tabularx}
{\rule{\temptablewidth}{1pt}}
\end{center}
\end{table}

\begin{table}[h!]
\tabcolsep 0pt \caption{Errors and timings of the 2D Coulomb potential in Example \ref{Coul2D} for anisotropic densities with $\sigma = 2$ computed on $[-12,12]\times \frac{1}{\gamma} [-12,12] $ with $h_x = 1/8, h_y = h_x/\gamma$ and $N = 192^2$. } \label{Tab_Coulomb2D_Aniso}
\begin{center}\vspace{-1em}
\def\temptablewidth{1\textwidth}
{\rule{\temptablewidth}{1pt}}
\begin{tabularx}{\temptablewidth}{@{\extracolsep{\fill}}p{1.05cm}cllll}
$\gamma$ & $\erh$ & $\|u\|_{\rm max}$ &$T_1$ &$T_2$ &$T_{total}$ \\ \hline  
$1$ &5.047E-16  & 1.773      & 1.00E-02 &  2.00E-03 &  1.20E-02\\
$2$ &5.479E-16  & 1.217      & 1.20E-02 &  3.00E-03 &  1.50E-02\\
$4$ &4.235E-16  & 7.902E-01  & 9.00E-03 &  2.00E-03 &  1.10E-02\\
$8$ &1.402E-15  & 4.902E-01  & 1.20E-02 &  2.00E-03 &  1.40E-02\\
$16$&8.387E-15  & 2.935E-01  & 1.20E-02 &  2.00E-03 &  1.40E-02\\
\end{tabularx}
{\rule{\temptablewidth}{1pt}}
\end{center}
\end{table}
%

\vspace{0.5cm}
\begin{exmp}\label{Poi2D}  The 2D Poisson potential. \end{exmp} 
For $\rho(\bx):= e^{-|\bx|^2/\sigma^2}=e^{-r^2/\sigma^2}$ with $r=|\bx|$ and $\sigma>0$, the 2D Poisson 
potential, with the kernel $U(\bx) = -\frac{1}{2\pi} \ln |\bx|$, can be obtained analytically as
\bea\label{eg-2d-Exact}
u(\bx)=\left\{\begin{array}{ll}-\frac{\sigma^2}{4}\,\left[{\textrm E}_1\left( \frac{|\bx|^2}{\sigma^2}\right)+2\ln(|\bx|)\right],
& \bx \neq \bzo ,\\[0.25cm]
\frac{\sigma^2}{4}\left(\gamma_e-\ln( \sigma^2)\right), &\bx = \bzo,
\end{array}
\right.
\eea
where ${\textrm E}_1(r):= \int_r^{\infty} t^{-1}e^{-t} {d}t$  for $r>0$
is the exponential integral function \cite{Handbook} and $\gamma_e\approx 0.5772156649015328606$ is the Euler-Mascheroni constant. 

The  2D Poisson  potential is computed on $ [-L,L]^2$ with mesh size $h_x = h_y$.
Table \ref{Tab_Poisson2D} shows the error $\erh$ and computation time with $\sigma = 1.2$ on $[-L,L]^2$ with different mesh sizes. Spectral accuracy and $O(N\ln N)$ efficiency can be observed from Tab.~\ref{Tab_Poisson2D}.

\begin{table}[h!]
\tabcolsep 0pt \caption{Errors and timings of the 2D Poisson potential in Example \ref{Poi2D} with $\sigma = 1.2$ on $ [-L,L] ^2$.} \label{Tab_Poisson2D}
\begin{center}\vspace{-1em}
\def\temptablewidth{1\textwidth}
{\rule{\temptablewidth}{1pt}}
\begin{tabularx}{\temptablewidth}{@{\extracolsep{\fill}}p{1.05cm}lllll}
$L =8$&$N$&  $\erh$ & $T_1$ & $T_2$ & $T_{total}$ \\ \hline
$h = 1$         & $16^2$       &3.768E-04 &0     &0     &0       \\
$h\!\! =\!\!1/2$& $32^2$       &3.331E-10 &1.00E-03 &0    &1.00E-03   \\
$h\!\! =\!\!1/4 $& $64^2$      &3.623E-15 &2.00E-03 &1.00E-03 &3.00E-03   \\
$h \!\! =\!\!1/8$& $128^2$     &2.988E-15 &6.00E-03 &1.00E-03 &7.00E-03   \\
$h \!\! =\!\!1\!/\!16$& $256^2$&5.085E-15 &2.30E-02 &4.00E-03 &2.70E-02   \\
\hline \hline
$L\!=\!16$&$N$&  $\erh$ & $T_1$ & $T_2$ & $T_{total}$ \\ \hline
$h = 1$         & $32^2$           &2.966E-04  & 1.00E-04 &         0   &1.00E-03  \\
$h\!\! =\!\!1/2$& $64^2$         &2.713E-10  & 2.00E-03 &         0   &2.00E-03  \\
$h\!\! =\!\!1/4 $&$128^2$       &3.856E-15  & 6.00E-03 &  2.00E-03   &8.00E-03  \\
$h \!\! =\!\!1/8$&$256^2$       &3.164E-15  & 2.60E-02 &  6.00E-03   &3.20E-02  \\
$h \!\! =\!\!1\!/\!16$& $512^2$&6.921E-15  & 2.47E-01 &  3.00E-02   &2.77E-01  \\
\end{tabularx}
{\rule{\temptablewidth}{1pt}}
\end{center}
\end{table}


\vspace{0.5cm}
\begin{exmp}\label{dipole-3d} The dipolar potential in 3D. \end{exmp} 
The 3D dipolar potential is defined by convolution as follows \cite{BC7,DipJCP,BaoJiangLeslie}:
\bea\label{dipolar3D:eq1}
 u(\bx) &= &-(\bn\cdot\bm)\,\rho(\bx) - 3\;\partial_{\bn\bm}
\left(\frac{1}{4\pi |\bx|} \ast \rho\right) \nonumber \\
&=& -(\bn\cdot\bm)\,\rho(\bx)-3\; \frac{1}{4\pi |\bx|} \ast \big(\partial_{\bn\bm}\, \rho\big) 
\eea
where $\bn, \bm$ are two given unit vectors in $\mathbb R^3$.
Note that the dipolar potential can actually be solved via the Coulomb potential by \eqref{dipolar3D:eq1} 
with the new source term
$\big(\partial_{\bn\bm} \rho\big)$.
 Numerically, the source term $\big(\partial_{\bn\bm} \rho\big)$ can be easily obtained  by differentiating the Fourier pseudospectral approximation of the fast decaying density.

Similarly, we consider a radial symmetric density $\rho(\bx)= e^{-|\bx|^{2}/\sigma^{2}}$, and the potential is given explicitly as 
\bea
 u(\bx) &= &-(\bn\cdot\bm)\,\rho(\bx) - 3\;\partial_{\bn\bm}
\left(\frac{1}{4\pi |\bx|} \ast \rho\right)\nonumber \\
&=&-(\bn\cdot\bm)\,\rho(\bx) - 3\; \partial_{\bn\bm} \left(\frac{\sigma^{2}\sqrt{\pi}}{4}  \frac{\text{Erf}(r/\sigma)}{r/\sigma}\right)\qquad \qquad  \nonumber \\
 &=& -(\bn\cdot\bm)\,\rho(\bx) - 3\;\bn^{T} \textbf{D} \;\bm,\eea
where $\delta_{ij}$ is the Dirac delta function and the Hessian matrix $\textbf{D}$ is given as follows
\beas
\textbf{D}_{ij}& =&\quad \delta_{ij} \left( \frac{\sigma^{2}}{2r^{2}} e^{-\frac{r^{2}}{\sigma^{2}}} - \frac{\sigma^{3}\sqrt{\pi}}{4 r^{3}}\text{Erf}
\left(\frac{r}{\sigma}\right) \right) +\\
&&\bx_{i}\bx_{j}  \left( -\frac{3 \;\sigma^{2}}{2\;r^{4}} e^{-\frac{r^{2}}{\sigma^{2}}} - \frac{1}{r^{2}} e^{-\frac{r^{2}}{\sigma^{2}}} + \frac{ 3\; \sigma^{3}\sqrt{\pi}} {4\;r^{5}} \text{Erf}\left(\frac{r}{\sigma}\right) \right),\quad \quad  i,j = 1,2,3.
\eeas

Table \ref{Tab_Dipolar3D} shows the error and timings of the 3D dipolar potential evaluation with $\sigma = 1.2$ and two randomly chosen vectors $\bn = (0.82778,0.41505,-0.37751)^T,
\bm = (0.3118,0.9378,-0.15214)^T$ on $[-8,8]^3$. Here, $T_{\rm pre}$ is the CPU time
for computing the source term $\partial_{\bn\bm} \rho$, $T_1,T_2$ and $T_{\rm total}$ are the 
same as those defined previously. We observe spectral accuracy and the timings show the expected scaling $O(N \ln N)$.

\begin{table}[h!]
\tabcolsep 0pt \caption{Errors and timings of the 3D dipolar potential 
in Example \ref{dipole-3d} with $\sigma = 1.2,\bn = (0.82778,0.41505,-0.37751)^T,
\bm = (0.3118,0.9378,-0.15214)^T$ on $[-8,8]^3$. } \label{Tab_Dipolar3D}
\begin{center}\vspace{-1em}
\def\temptablewidth{1\textwidth}
{\rule{\temptablewidth}{1pt}}
\begin{tabularx}{\temptablewidth}{@{\extracolsep{\fill}}p{1.05cm}clllllll}
$L =8$&$N$&  $\erh$ &$T_{\rm pre}$ &$T_1$ & $T_2$ & $T_{total}$ \\ \hline
$h = 1$         & $16^3$  &1.380E-02 &          0   & 2.00E-03  &       0    & 2.00E-03 \\
$h\!\! =\!\!1/2$& $32^3$  &2.647E-07 &  2.00E-03  & 1.50E-02  & 2.00E-03   &1.90E-02  \\
$h\!\! =\!\!1/4 $& $64^3$ &1.430E-14 &  1.70E-02  & 2.00E-01  & 1.90E-02   &2.35E-01  \\
$h \!\! =\!\!1/8$& $128^3$&4.076E-14 &  1.96E-01  & 1.68  & 2.20E-01   &2.10  \\
\end{tabularx}
{\rule{\temptablewidth}{1pt}}
\end{center}
\end{table}

\vspace{0.5cm}
\begin{exmp}\label{ds} The Davey-Stewartson (DS) nonlocal potential. \end{exmp} 
In the DS equation, the nonlocal potential can be given by a convolution as follows:
\bea
u(\bx) = -\frac{1}{2\pi} \ln |\bx| \ast (\partial_{xx}\rho),\quad \quad  \bx \in \mathbb R^2.
\eea
For a Gaussian density $\rho(x,y) = \pi \;e^{-\pi^2 (x^2+y^2)}$,  the DS nonlocal potential is given explicitly,  in polar coordinates, as
\bea
\Phi(r,\theta)=  -\left(\frac{\pi}{2}\;e^{-\pi ^2 r^2} +\cos(2 \theta )\;  e^{-\pi ^2 r^2} (2\pi r^2)^{-1}(1+\pi ^2 r^2-e^{\pi ^2 r^2})\right).
\eea
Table \ref{Tab_DS} displays the error and computational time of the DS nonlocal potential on $[-8,8]^2$. 
We observe the scaling $O(N\ln N)$ and rapid decrease of the error as the mesh size gets smaller, although the best reached precision is below those of the previous 2D Poisson/Coulomb examples. 
In this context, note that the parameter $\sigma$ is larger compared to the preceding test in Example~\ref{Poi2D}, i.e., a finer resolution would be needed for the more localized density.  

\begin{table}[h!]
\tabcolsep 0pt \caption{Errors and timings of the DS nonlocal potential in Example \ref{ds} on $ [-8,8] ^2$.} \label{Tab_DS}
\begin{center}\vspace{-1em}
\def\temptablewidth{1\textwidth}
{\rule{\temptablewidth}{1pt}}
\begin{tabularx}{\temptablewidth}{@{\extracolsep{\fill}}p{1.05cm}lllll}
$L = 8$ &$N$&  $\erh$ & $T_1$ & $T_2$ & $T_{total}$ \\ \hline
$h\!\! =\!\!1/2$& $32^2$        &1.474      &1.00E-03 &  0         &1.00E-03  \\
$h\!\! =\!\!1/4 $& $64^2$       &5.720E-03  &2.00E-03 &  0         &2.00E-03  \\
$h \!\! =\!\!1/8$& $128^2$      &3.974E-09  &5.00E-03 &  2.00E-04  &6.00E-03  \\
$h \!\! =\!\!1\!/\!16$& $256^2$ &4.536E-13  &2.20E-02 &  7.00E-03  &2.80E-02  \\
\end{tabularx}
{\rule{\temptablewidth}{1pt}}
\end{center}
\end{table}

\section{Conclusions}
Starting from the convolution definition, we presented an efficient and accurate algorithm for a class of nonlocal (long-range) potentials, where the densities are assumed to be smooth and fast decaying. The method use a Gaussian-sum approximation of the singular convolution kernel to split the convolution into two integrals, namely a long-range
regular integral and a short-range singular integral.  
Due to the high-resolution GS approximation obtained with sinc quadrature, the regular 
integral was computed with FFT and the singular integral evaluation was done with a low-order Taylor expansion of the density. 
The algorithm achieves spectral accuracy and is essentially as efficient as FFT algorithms with a computational complexity at $O(N \ln N)$, where $N$ is the total number of points in the discretization of physical space. 
The method was implemented in Fortran and verified for several different potentials, including the 2D/3D Coulomb potential, the 2D Poisson,
the 3D dipolar potential and the  Davey-Stewartson nonlocal potential.

The algorithm is suitable for parallel computation, e.g., MPI or GPU parallel computation and the development of parallel version is on-going.
We shall mention here that the algorithm could possibly be adapted to the magnetostatic potential in stray field computations for micromagnetic simulations \cite{exl_2014,exl_2012}. 


\section*{Acknowledgments}
We acknowledge financial support by the Austrian Science Fund (FWF) SFB ViCoM (F4112-N13) (L. Exl, N. Mauser),  
grant No. I830 (project LODIQUAS) and the Austrian Ministry of Science
and Research via its grant for the WPI (N. Mauser, Y. Zhang).  
The computation results presented have been achieved by using the Vienna Scientific Cluster (VSC).


\end{document}